# FISH DETECTION USING MORPHOLOGICAL APPROACH BASED-ON K-MEANS SEGMENTATION

**Shoffan Saifullah[1\*)], Andiko Putro Suryotomo[2], Bambang Yuwono[3]**

[1, 2, 3]Department of Informatics Engineering, Universitas Pembangunan Nasional Veteran Yogyakarta

[\*)]E-mail: shoffans@upnyk.ac.id

### ABSTRACT

Image segmentation is a concept that is often used for object detection. This detection has difficulty detecting objects with backgrounds that have many colors and even have a color similar to the object being detected. This study aims to detect fish using segmentation, namely segmenting fish images using k-means clustering. The segmentation process is processed by improving the image first. The initial process is preprocessing to improve the image. Preprocessing is done twice, before segmentation using k-means and after. Preprocessing stage 1 using resize and reshape. Whereas after k-means is the contrast-limited adaptive histogram equalization. Preprocessing results are segmented using k-means clustering. The K-means concept classifies images using segments between the object and the background (using k = 8). The final step is the morphological process with open and close operations to obtain fish contours using black and white images based on grayscale images from color images. Based on the experimental results, the process can run well, with the ssim value close to 1, which means that image information does not change. Processed objects provide a clear picture of fish objects so that this k-means segmentation can help detect fish objects.

## 1. INTRODUCTION

Object detection can be done using the segmentation method (Priyadharsini & Sharmila, 2019). This detection can differentiate between objects and backgrounds (Saifullah, 2020c; Yudhana, Sunardi, & Saifullah, 2016, 2017). However, the segmentation process requires some initial or preprocessing methods to provide clarity to the processed image. This preprocessing helps to improve the image for further processing. This process is needed because the base of the image acquired has imperfections, for example, blur, broken, too light/dark, and so on, especially in detecting fish, which objects are below the surface of the water. The object must have a different image from the original. So it is necessary to do preprocessing.

The process of detecting fish objects is mostly carried out by several methods such as deep learning (Salman et al., 2020; Sung, Yu, & Girdhar, 2017; L. Yang et al., 2020), probabilistic modelling (Salman, Maqbool, Khan, Jalal, & Shafait, 2019), segmentation (Garcia et al., 2020; Ibrahim, Ahmed, Hussein, & Hassanien, 2018; Sheng et al., 2016). One of the references focuses used in this study is fish image segmentation using k-means clustering and morphological development (Saifullah, 2020d; Sunardi, Yudhana, & Saifullah, 2017; Yao, Duan, Li, & Wang, 2013). This study's difference is the image preprocessing part which is used, namely using image enhancement with the contrast limited adaptive histogram equalization (CLAHE) and HE method.

Besides, other studies apply the k-means clustering algorithm and otsu (Sheng et al., 2016) maupun l\*a\*b color space (Saifullah, 2020b; Tran et al., 2018) in the process of color image segmentation on fish objects. The koi fish classification based on the HSV color space concept was also carried out in this study (Kartika & Herumurti, 2016). Other methods are also implemented in fish image segmentation, such as the salp swarm algorithm (Ibrahim et al., 2018). So that the research conducted is different from the method used in previous studies.

This article discusses the topics related to the introduction in the first part regarding the introduction, problems, and related research (gaps). The second part discusses the methods used in this research, from preprocessing to morphology and edge detection. Furthermore, the results and discussion of this study are described and discussed in the third section. Last is the conclusion of the research conducted.

## 2. METHOD

This research method uses the concept of morphology and segmentation approaches. These two concepts complement each other for better results. In this study, Morphology is used to improve segmentation results. The details of the method used are shown in Figure 1.

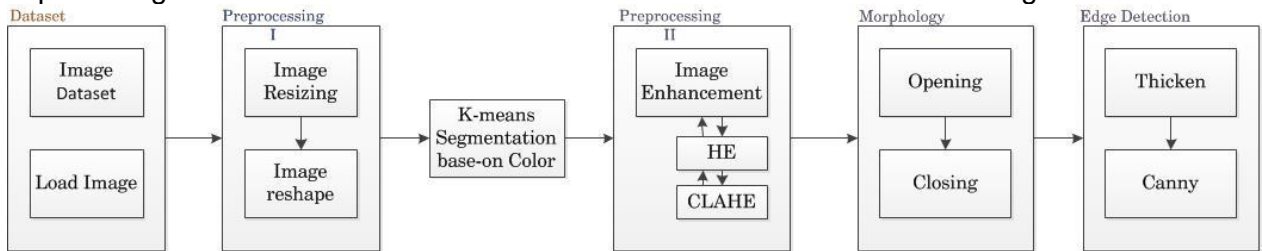

Figure 1. Fish Detection Process Using Morphology and Edge Detection

Figure 1 represents the steps taken in this study, starting from using a fish image dataset to the edge detection process. This dataset is a secondary dataset taken from (Yao et al., 2013) and QUT_fish_data (source: https://wiki.qut.edu.au/display/raq/Fish+Dataset). Fish images are processed using several preprocessing methods, segmentation, morphology, and edge detection.

### 2.1 Image preprocessing

Image processing in this research is carried out in two stages. The first stage is carried out by using two processes, namely resizing and reshaping. This resizing aims to change the size of the image into pixels. In the research, the size is changed to be smaller than the original image size (Saifullah, 2020a), aiming for image processing at the time of segmentation being faster and not changing the information contained in it. Meanwhile, reshape is used to modify the dimensions of the original matrix to be as desired. This process is a continuation of resizing, and the results use a size that matches the size of the resize.

Preprocessing Phase II is carried out after the color image segmentation process using K-means clustering. The purpose of this preprocessing stage is to improve image quality by using the repair histogram. Histogram repair using two methods, namely Histogram Equalization (HE) and Contrast Limited Adaptive Histogram Equalization (CLAHE) based on grayscale images (conversion of color images (RGB) to grayscale using (1)) (Putra, Supriadi, Wibawa, Pranolo, & Gaffar, 2020; Saifullah, 2020c; Saifullah, Sunardi, & Yudhana, 2017). HE provides an increase in image quality based on the histogram's equalization so that the resulting histogram has a more even distribution than the histogram of the original image. The improvement of the HE method uses (2).
Meanwhile, the next process is CLAHE, which is the development of HE. CLAHE has the concept of HE with the maximum height limit value of the histogram. CLAHE calculation uses the clip limit concept as a histogram constraint (3).

$$gray = 0,2989 * R + 0,587 * G + 0,1141 * B \qquad (1)$$

$$hi = \frac{n_i}{n}, i = 0, 1, 2, \dots, L-1 \qquad (2)$$

$$\beta = \frac{M}{n}(1 + \frac{\alpha}{100}(s_{max} - 1)) \qquad (3)$$

The color image needs to be converted into a grayscale image using (1). Each color (R = Red, G = Green, B = Blue) is performed with a conventional formula constant. Then it will be processed with Image Enhancement (HE and CLAHE). HE has L as the degree of gray. ni is the number of pixels of gray i, and n is the number of all pixels. Besides that, CLAHE has M, which is the size of the region. Besides, N is a grayscale value. The a is a clip factor as the addition of the histogram's limit, which is between 0-100.

Between the two preprocessing, there is a segmentation process. This segmentation aims to label the pixels in the image. The segmented image can be easily analyzed because it has been represented in several significant parts (Jamal, Manaa, Rabee'a, & Khalaf, 2015). Image segmentation can be processed using edge extraction, threshold, region, and others (H. Y. Yang, Zhao, Xu, & Liu, 2018). Image segmentation in this study uses the k-means clustering method and ends with edge detection.

K-means segmentation in images is used with the block modeling concept that divides the image area into several parts. K-means is one of the unsupervised learning methods (Caron, Bojanowski, Mairal, & Joulin, 2019; Nguyen & De Baets, 2019) which accepts input in the form of labelless data. This method is prevalent, simple, and fast in the process (Garcia-Dias, Vieira, Lopez Pinaya, & Mechelli, 2020). K-means is processed using five main steps, as shown in Figure 2.

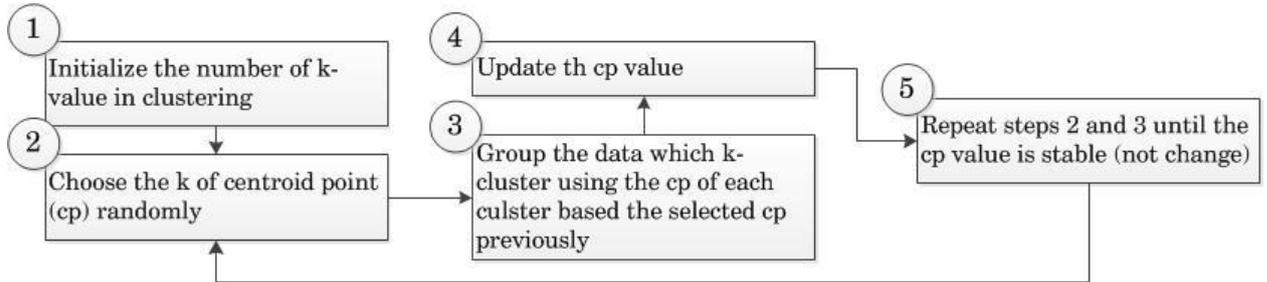

Figure 2. Steps of K-means Algorithm in Making Data Clusters

The K-means process involves grouping the data as input and creating class targets. The incoming data is in the form of data or objects with the k-cluster. The clustering process using this method uses the calculation of the closest data distance to the centroid. The distance used is the euclidean distance (Loohach & Garg, 2012), which calculates two adjacent data using (4) and (5).

$$d(i,j) = \sqrt{\sum_{k=1}^{p}(x_{ik} - x_{jk})^2} \qquad (4)$$

$$\mu_k = \frac{1}{N}\sum_{q=1}^{N_k} x_q \qquad (5)$$

The process of calculating the distance with the euclidean distance (4) with the centroid point (5) is based on the data. The *d(i,j)* is the distance of object i to j on p (data dimension) with each object's coordinates (*xij* or *xjk*). The update uses the *μk* (centroid point of the k-cluster) where *Nk* is many data and *xq* is the q-data in the k-cluster.

Besides, the process of improving the segmentation results is carried out using morphological operations. This operation can change the shape of the original image based on the operator used. This operation can be performed on grayscale or black and white images. In this study, the operations used include opening, closing. Opening the image is done using erosion to reduce or erode the edges of the object, followed by dilation (enlargement of the object segment). While closing is the opposite of the opening method (dilation-erosion). Closing aims to close small holes and join adjacent objects. Dilation operations are performed by adding layers to each object (thickening the object). The dilation application uses a structure element (strel) line with a length of 1 and 45 degrees.

Another morphology is edge detection with "thicken" and "canny" operations. Thicken is used to thicken selected areas of the foreground pixels of a black and white image. In the concept of edge detection, detection is processed by certain operations to detect homogeneous images with different brightness levels. Meanwhile, the canny operator is the optimal edge detection using the Gaussian Derivative Kernel concept. Canny detection results have smooth edges because they are filtered based on the initial image. This operator is carried out by several processes such as grayscaling, Gaussian filters, intensity gradients, Non-Maximum Suppression, Double Thresholding, Edge Tracking by Hysteresis, and Cleaning Up. This detection's result is to show that fish objects can be detected based on the edges.

The final result of this detection is tested using the structural similarity index method (SSIM). This method is used to express perceptions that transform the image into a reference in structural information. This method's measurement is based on several components, namely luminance, construction, and structure, with a comparison of two images of the same size. Calculations can be done using (6).

$$SSIM(x, y) = [l(x, y)]^\alpha [c(x, y)]^\beta [s(x, y)]^\gamma \tag{6}$$

Each of these components (l = luminance, c = contrast, and s = structure) compares two images with α, β, and γ being positive constants (Kumar & Moyal, 2013). This calculation has a value between 0 and 1, which means that the smaller the value shows a large difference between the resulting image and the original image/different information (Leksono, Raharjo, & Safitri, 2019).

## 3. RESULTS AND DISCUSSION

In this study, an analysis and experiment process was carried out regarding the process of fish detection. The process is carried out based on the dataset used in this study, namely the fish's image, then preprocessing, segmentation, and morphology are carried out. The process is sequential, as seen in Figure 1. The details of the results and discussion are presented in this section.

### 3.1 Fish Image Dataset

The dataset used in this study is a secondary dataset taken from other research on fish images. The sample data used in this study are shown in Figure 3.

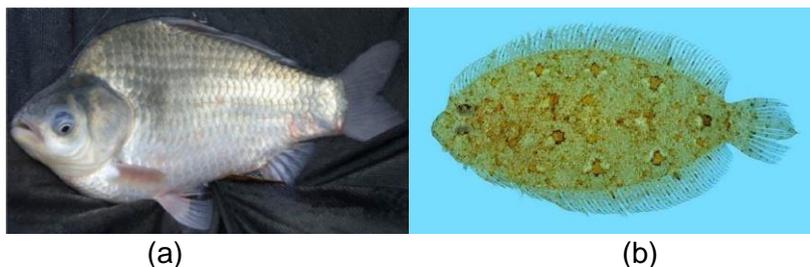

(a)                                    (b)

Figure 3. Fish samples used in this study (a) (Yao et al., 2013)  and (b) QUT_fish_data

### 3.2 Preprocessing Phase I

Image preprocessing stage I is carried out using two methods, namely, resize and reshape. As in Figure 3. (a), the original image has a size of 243x434 pixels. This image, when processed, requires a long time and lengthy process. So it is done resizing and reshaping the image to speed up the process. Resizing is done by reducing the size to 50x50 pixels for each color (3 colors = RGB). The reshaping follows the resize size to obtain numerical data with a size of 2500x3, which shows the multiplication of the resized image size and the color components. The sample used comes from 2 datasets. The second sample is from QUT_fish_data, where each dataset used has different sizes, so it needs to be resized.

### 3.3 K-means Segmentation Based-on Color Image

Based on Sample Figure 3. (a), the segmentation process is carried out using the k-means clustering method to obtain a segmented image. This process uses the resized image to make the size smaller and blur, as in Figure 4 (a). Figure 4 (b) processed segmentation using K-means clustering with a value of k = 8.

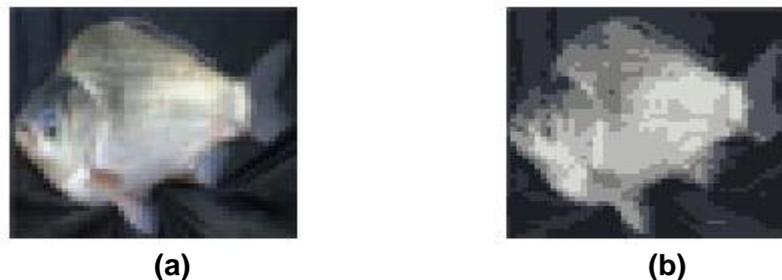

**(a)**                                        **(b)**

Figure 4. (a) Image Resized and Image Result of K-means Clustering Segmentation

The segmentation process with clustering is done with experiments with several values of k, ranging from 2 to 9. The best result of this segmentation is to use the value of k = 8, where all the image segmentation areas are covered. When implemented less than 8, a trimmed fish image area will not be seen ultimately. Meanwhile, if the k value is more than 8, the image has additional areas when processed until the end, so that the additional area needs to be removed.

### 3.4 Preprocessing Phase II

Based on the image processed by segmentation (Figure 3), the resulting histogram is as in Figure 5. Figure 5. (a) shows the distribution of the combined values of 3 RGB colors depicted in one histogram. Figure 5. (b) shows the histogram of the reshape image with a value distribution of 0 to 1 and maximum intensity of 150.

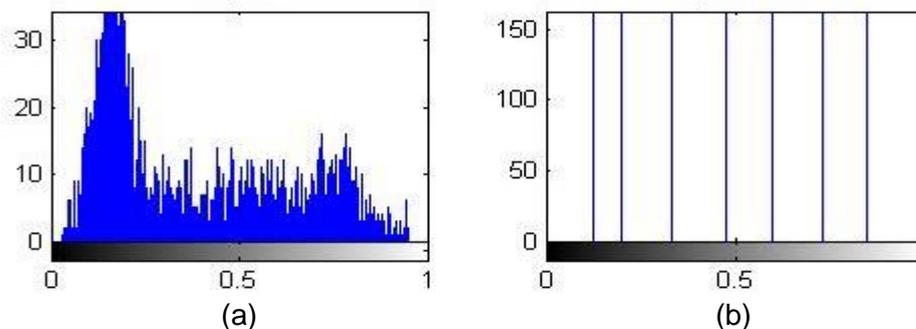

(a)                                        (b)

Figure 5. (a) Combined Color Histogram from the Resized Image and (b) Histogram of the Segmented Image Reshape Result

Based on Figure 5, the second preprocessing is to repair the image using the histogram concept. Repairs were carried out using a combination of the HE and CLAHE methods. This combination can provide a clearer and easier image for the next process (Saifullah, 2019).

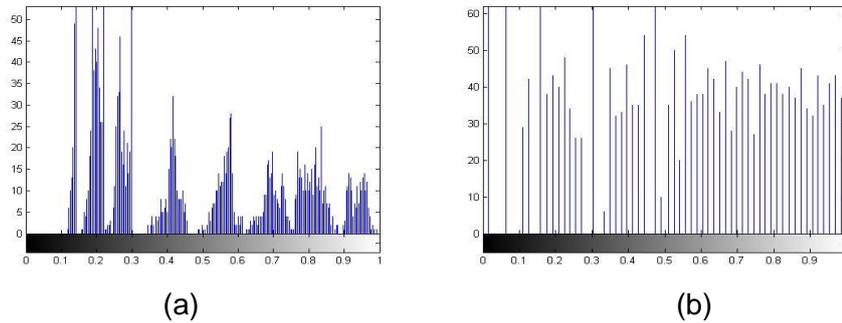

(a)                                     (b)

Figure 6. (a) CLAHE Processing Results, and (b) Continued using HE

The histogram repair is carried out based on the histogram shown in Figure 5, whose distribution follows the original image. Figure 5 shows that the histogram distribution is generally uneven, and some are very high, and some are very low. So that in this study, implementing image quality improvements using CLAHE and HE. The CLAHE results in the histogram look like in Figure 6. (a) and still have an uneven distribution, so that further processing with HE (HE results are shown in Figure 6 (b). The results of the combination of the two methods provide an average histogram whose range is less Figure 6 shows the results of each image shown in Figure 7. (a) and (b). These results are converted into black and white (binary) images for morphological processes. The binary image results are shown in Figure 7. (c) Binary image has only two colors, namely 0 and 255. The value 0 represents black, and 255 represents white/light.

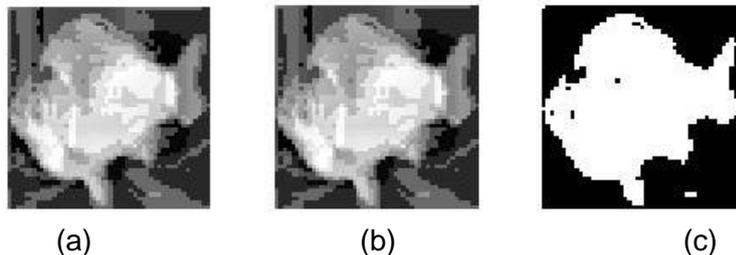

(a)                          (b)                          (c)

Figure 7. Image Result of Histogram Enhancement with (a) CLAHE and (b) Advanced HE, and (c) Image Result of Conversion (Binary Image/Black-White)

### 3.5 Morphology and Edge Detection

This research ends with a morphological process and edge detection. Morphology was carried out using opening, closing, and thicken to obtain an image as in Figure 8. (a). Figure 8. (a) is the result obtained from Figure 7. (c), processed by morphology. The thickening process provides adequate space and can be seen in Figure 8. (a). The final process using edge detection with the Canny method produces an image, as shown in Figure 8. (b). The final image shows a fish image pattern obtained using edge detection. When implemented in the original image, it is obtained as in Figure 8. (C). Another example is shown in Figure 8. (d). The results of this detection still have shortcomings; if the background is close to the fish's color, the fish will be detected.

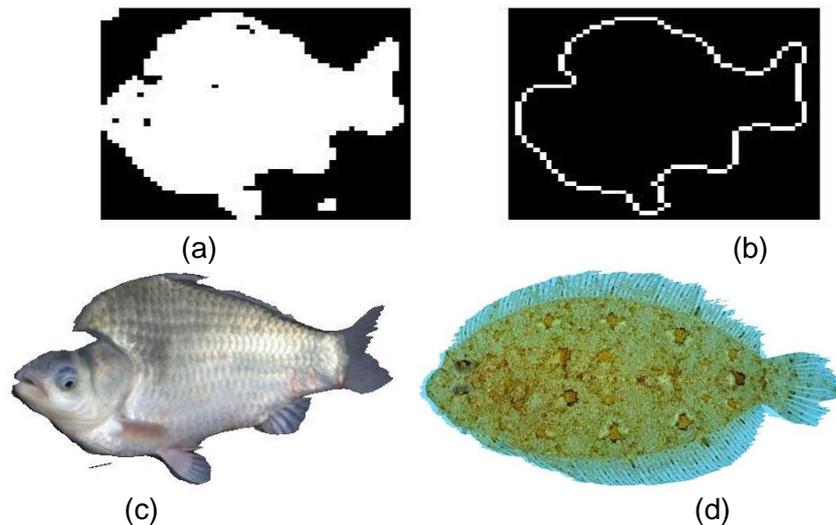

(a)　　　　　　　　　　　　　(b)

(c)　　　　　　　　　　　　　(d)

Figure 8. Image result (a). Morphology, (b). Edge Detection, and (c). Segmentation of Fish Image in Dataset 1 and Example of Segmentation Results in Dataset 2

### 3.6 Perhitungan Pengunjian dengan SSIM

Based on the resulting segmentation results, SSIM is calculated to check the results' similarity with the original image. The SSIM results are shown in Figure 9. The SSIM value has a value distribution between 0.9965 to 1. The SSIM average value of the distribution is 0.9994. The similarity generated from this SSIM calculation is close to the value of 1, so it can be said that the information generated is following the original image and can be used for fish detection/segmentation.

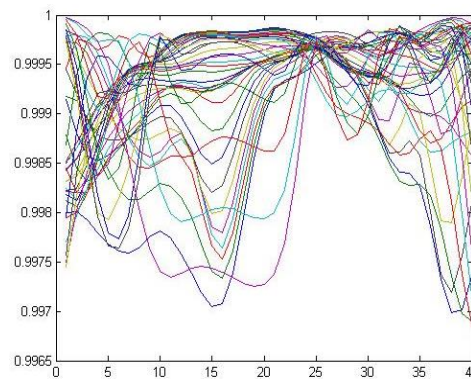

Figure 9. Distribution of SSIM values based on the Result Image and the Original Image

## 4. CONCLUSION

This study's segmentation uses the k-means method with k = 8 based on color images (RGB). The initial process is only to use resizing and reshaping to speed up the segmentation process. The segmentation results on the color image are not yet evident, so it must be processed using advanced preprocessing with histogram correction. As a result, the second preprocessing is processed with morphology and edge detection. The object details can be seen clearly and can be divided between the fish and the background. The final segmentation image corresponds to the image being processed, and the detected object can be seen clearly. Based on the final image, the SSIM value distribution shows the value is almost close to 1, namely 0.9994. This result means that the resulting image has the complete information from the original image and can be used for segmentation.

## 5. AKNOWLEDGEMENT

Thanks to the authors, the Department of Informatics, Yogyakarta Veteran National Development University has supported this research publication. Besides, it is also conveyed to the team of lecturers who have collaborated to complete this article.